\documentclass[iop,apj,numberappendix,appendixfloats]{emulateapj}
\usepackage{graphicx}
\pdfoutput=1

\usepackage{natbib}

\LongTables
\def \lyaf {Ly$\alpha$ forest}
\def \Lya  {Ly$\alpha$}
\def \lya  {Ly$\alpha$}

\def \lyb  {Ly$\beta$}

\def \ly5  {Ly-5}
\def \ly6  {Ly-6}
\def \ly7  {Ly-7}

\def \lnhi {$\log N_{\rm HI}$}

\def \lyaf {Lyman--$\alpha$ forest}

\newcommand{\kms}{${\rm km\,s}^{-1}$}
\newcommand{\HI}{\ion{H}{1}}

\def \nqsos {300}
\def \nspectra {831}
\def \ntotspectra {1577}

\submitted{Submitted to the Astronomical Journal}
\shortauthors{O'Meara et al.}
\shorttitle{KODIAQ DR2}

\begin{document}

\title{The Second Data Release of the KODIAQ Survey}

\author{
J.M. O'Meara\altaffilmark{1},
N. Lehner\altaffilmark{2},
J.C. Howk\altaffilmark{2},
J.X. Prochaska\altaffilmark{3},
A.J. Fox\altaffilmark{4},
M.S. Peeples\altaffilmark{4,5},
J. Tumlinson\altaffilmark{4,5},
B.W. O'Shea\altaffilmark{6}
}
\altaffiltext{1}{Department of Chemistry and Physics, Saint Michael's College, One Winooski Park, Colchester, VT 05439}
\altaffiltext{2}{Department of Physics, University of Notre Dame, 225 Nieuwland Science Hall, Notre Dame, IN 46556}
\altaffiltext{3}{University of California/Lick Observatory, Santa Cruz, 1156 High Street, CA 95064}
\altaffiltext{4}{Space Telescope Science Institute, 3700 San Martin
  Drive, Baltimore, MD, 21218}
\altaffiltext{5}{Department of Physics and Astronomy, Johns Hopkins University, Baltimore, MD 21218}
\altaffiltext{6}{Department of Computational Mathematics, Science and Engineering, Michigan State University, East Lansing, MI 48824}

\begin{abstract}
  We present and make publicly available the second data release (DR2)
  of the Keck Observatory Database of Ionized Absorption toward
  Quasars (KODIAQ) survey.  KODIAQ DR2 consists of a
  fully-reduced sample of \nqsos\ quasars at $0.07 < z_{\rm em} <
  5.29$ observed with HIRES at high resolution ($36,000 \le R \le
  103,000$).  DR2 contains \nspectra\ spectra
  available in continuum normalized form, representing a sum total
  exposure time of $\sim 4.9$ megaseconds on source.  These co-added spectra
  arise from a total of \ntotspectra\ individual exposures of quasars
  taken from the Keck Observatory Archive (KOA) in raw form and
  uniformly processed.  DR2 extends DR1 by adding $130$ new quasars to the sample,
  including additional observations of QSOs in DR1.  All new data in DR2
  were obtained with the single-chip Tektronix TK2048 CCD configuration 
  of HIRES 
  in operation between 1995 and 2004.  DR2 is publicly available
  to the community, housed as a higher level science product at the
  KOA and in the \textit{igmspec} database (v03).  
\end{abstract}

\keywords{absorption lines -- intergalactic medium -- Lyman limit
  systems -- damped Lyman alpha systems}

\section{Introduction}\label{intro}

The High-Resolution Echelle Spectrograph \citep[HIRES]{vogt94} on the Keck I telescope has been a spectacular instrument for the study of intergalactic and circumgalactic gas around galaxies since its introduction in 1994.  The large aperture of Keck I coupled with the high spectral resolution of HIRES ($R \approx 35,000 - 100,000$) gave rise to a 20 year golden age of QSO absorption line studies. HIRES has allowed the best determinations of the primordial D/H ratio \citep{burles98,omeara01,cooke14}, a test of Big Bang nucleosynthesis and fundamental physics,  and it provided high precision constraints on  the fine-structure constant \citep{murphy01,murphy03}.  HIRES has been instrumental in studying the physics of the intergalactic medium (IGM) and thereby cosmology \citep{rauch97, simcoe04, tytler07, bolton14}, including the highest redshifts accessible at the end of reionization \citep{bolton10, becker11}. HIRES has played leading roles in the determination of the metallicity distribution in and physics of damped \Lya\ systems \citep{wolfe05, rafelski12}. HIRES has also been critical in the study of circumgalactic gas at all redshifts \citep{rudie12, lehner13, werk13, wotta16}, 
including in setting the scale for the Milky Way's high velocity cloud population \citep{thom08, wakker07, wakker08, smoker11}.

As part of their contribution to the Keck Observatory, NASA has provided  an archival service to give access to the data from HIRES (and now the other Keck instruments as well). The Keck Observatory Archive (KOA\footnote{{\tt http://www2.keck.hawaii.edu/koa/public/koa.php}} holds the vast majority of all HIRES observations of distant QSOs. These data are publicly accessible to all interested researchers. However, they are nominally only available in their raw form. 
Given the idiosyncrasies of data collection  by a large number of observers operating under disparate conditions, there is a significant hurdle to using these data directly for science. This  is especially true given the difficulties in coaddition, a process requiring an order-by-order continuum placement for optimal results. 

In order to address these issues, we have undertaken a NASA-funded processing of all of the extant HIRES data for the study of QSO absorption lines to provide fully-reduced QSO spectra to the community. This first data release (DR1) of our Keck Observatory Database of Ionized Absorption toward Quasars (KODIAQ) was presented in \cite{omeara15}, which included spectra for 170 QSOs.  The original science goal of the KODIAQ survey has been to study of the \ion{O}{6} absorption in strong HI absorbers at high $z>2.2$ \citep{lehner14}, revealing unique properties of these absorbers. Besides our original program  the data making up DR1 have motivated our new KODIAQ Z survey aimed to determine the metallicity of the strong HI absorbers ($15 < $\lnhi$<19$) at $z>2$, and have been used in a number of new surveys.
These include the characterization and interpretation of the small-scale structure  in the \Lya\ forest \citep{rorai17}, the metallicity distribution of  of strong \HI\ systems at intermediate- and high-redshifts \citep{fumagalli16,lehner13,lehner16,wotta16}, the statistical characterization of \ion{Mg}{2} and LLS absorbers \citep{mathes17,prochaska15}, characterization of extremely low metallicity gas, including for the determination of primordial D/H  \citep{cooke16,crighton16}, and even the study of the extended reaches of extremely low-redshift galaxies \citep{dutta16}. 

The DR1 datasets were confined to those taken after the HIRES upgrade to a three-CCD mosaic in 2004 due to the shift in data reduction approach associated with that upgrade.  This excluded the first decade of HIRES data, including many exceptional datasets.  Here we present KODIAQ DR2, which now includes HIRES spectra of QSOs from the years 1995--2004.  Our paper lays out our data reduction approach, which varies somewhat from that discussed in \cite{omeara15} due to the difference in detectors, and presents the statistics of the combined DR1+DR2 datasets.  Together these releases contain $300$ QSOs covering the redshift range $0.07 \la z_{em} \la 5.29$. All of the data presented here are now available through the KOA and will be available with the distribution of  v03 of the {\tt specdb} dataset \footnote{http://github.com/specdb} \citep{prochaska17}.

\section{The Data}\label{data}
A full discussion of HIRES and its data, including the process
of downloading and ingesting the raw data into the KODIAQ database is
presented in \cite{omeara15}.  The new data presented here in DR2 all
stem from HIRES observations by multiple PIs between 1995 and 2004.  Table 1
presents the HIRES deckers used across DR2 and their corresponding
spectral resolution.  As with DR1, the majority of observations were
made with the C1 or C5 decker providing $\approx 6$ and $\approx 8 \, \rm km \, s^{-1}$
FWHM resolution respectively. 

Data obtained prior to the HIRES upgrade in 2004 had slightly reduced
total spectral coverage due to the size of the detector.  As a result,
observations were frequently made in matched sets where the echelle
angle was kept fixed, but the cross-disperser angle varied so as to
provide fewer spectral gaps, particularly at redder wavelengths.  The
overall throughput of the pre-2004 detector was lower, resulting in a
longer average per-exposure integration time as compared to post-2004
upgrade observations.  

\subsection{Data Reduction}\label{redux}
The data reduction largely followed the same procedures as outlined in
\cite{omeara15}.  However, for the new data in DR2, only data in the
single-chip configuration was reduced, whereas in \cite{omeara15}, only
three-chip mosaic data was involved.  Minor differences in the data
reduction with HIRedux data reduction package\footnote{http://www.ucolick.org/$\sim$xavier/HIRedux/}
arise from the change in detectors.  First, the
single-chip detector has a blemish known as the ``ink spot'' in the
center of the detector.  
The pixels associated with the ink spot were masked in the 
extracted spectrum and ignored in any further processing 
so as not to confuse it with real QSO absorption lines.
Second, the flat fielding procedure from \cite{omeara15} was not
applied as the instrument was changed in 2004, thus negating the
ability to construct the nominal HIRedux pixel flats.  Instead, we use the internal flats both for order edge identification and pixel flat-fielding.

Finally, multiple spectra showed significant echelle order overlaps in their
bluest orders, Typically at wavelengths $\lambda < 3800$\AA, significantly complicating sky subtraction, source
identification, and source extraction.  In most cases, these orders
were trimmed off.  
 Therefore, the spectral coverage of these objects
is smaller than is quoted in the KOA.  It is also worth noting that
the pre-2004 hires detector was far less blue sensitive than its
successor.  As a result, observations at very blue ($\lambda < \sim
3400$\AA) wavelengths often had very long exposure times, some in
excess of two hours per integration.  For these exposures, unless a
number of other exposures of the same object were taken with the same
setup, cosmic ray rejection will likely not be optimal.
As with the DR1, continua were fit on an order-by-order basis in the
manner described in \cite{omeara15}.  The full DR2 sample contains
over $15000$ echelle orders that have been continuum fit.

\subsection{KODIAQ at the KOA and \it{igmspec}}\label{koa}
As with DR1, the DR2 data products are available for community
download from the KOA.  DR2 supersedes DR1 at the KOA and contains the
full 300 quasar sample.  As with DR1, users may search for and
download individual DR2 quasar sight line flux and error spectra, or
the full DR2 sample all at once.  Machine readable tables and files
exist for each spectrum to link back to the raw KOA data as needed.
DR2 also makes available all intermediate data reduction steps grouped
on an observing run by
observing run basis.  Spectra from DR2 are also available in the v03 release of the
\textit{igmspec} database (see \cite{prochaska17} for details).

\section{Properties of KODIAQ DR2}\label{dr2}
The KODIAQ DR2 comprises HIRES observations of 300 quasar lines of
sight in total.  Of these, 130 quasar sight lines are new since DR1,
along with many new additional observations of some of the DR1 quasars.
Table 2 presents the new data since DR1.  As in \cite{omeara15},
quasars are named according to their J2000 R.A./Decl. coordinates as
resolved by either SIMBAD or SDSS.  Many quasars were observed by
multiple PIs in multiple instrument configurations.  For a given
PI+configuration, the data were combined.  Table 2 lists the exposure
time of the combined spectra, including the wavelength coverage of the
spectra after trimming poorly extracted orders when necessary (see above).  We note that these exposure times are only a crude measure of data quality, as they do not include the observing conditions.

Table 3 presents the full DR2 sample of 300 quasars.  The rest
wavelengths listed in Table 3 are given at the quasar emission
redshift, i.e  $\lambda_{\rm{r}} = \lambda_{\rm{obs}} / (1.0 +
z_{\rm{em}})$.  
The full DR2 comprises 1577 individual exposures,
grouped into 831 spectral co-adds.  The aggregate exposure time of the
full DR2 sample is $\sim 4.9$ megaseconds.  DR2 represents a
significant increase over DR1, which was comprised of 170 quasar
sight lines, 240 co-additions, 567 individual exposures, and 1.6
megaseconds of total exposure time.

\subsection{General Properties}\label{general}
\begin{figure*}[ht]
\epsscale{0.7} 
\plotone{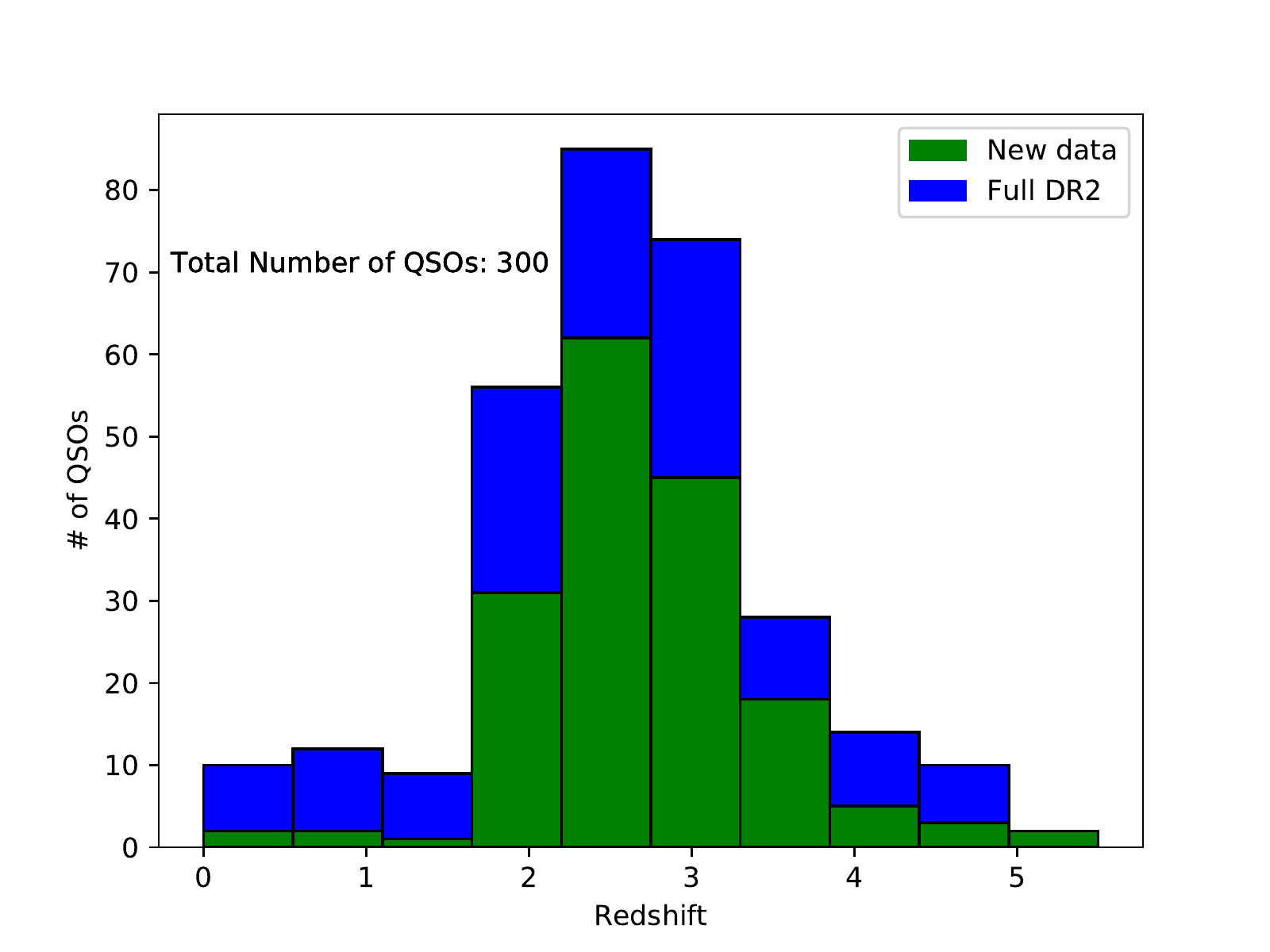}  
\caption{Redshift distribution of KODIAQ DR2 quasars. \label{fig_zhist}}
\end{figure*}

\begin{figure*}[ht]
\epsscale{0.7} 
\plotone{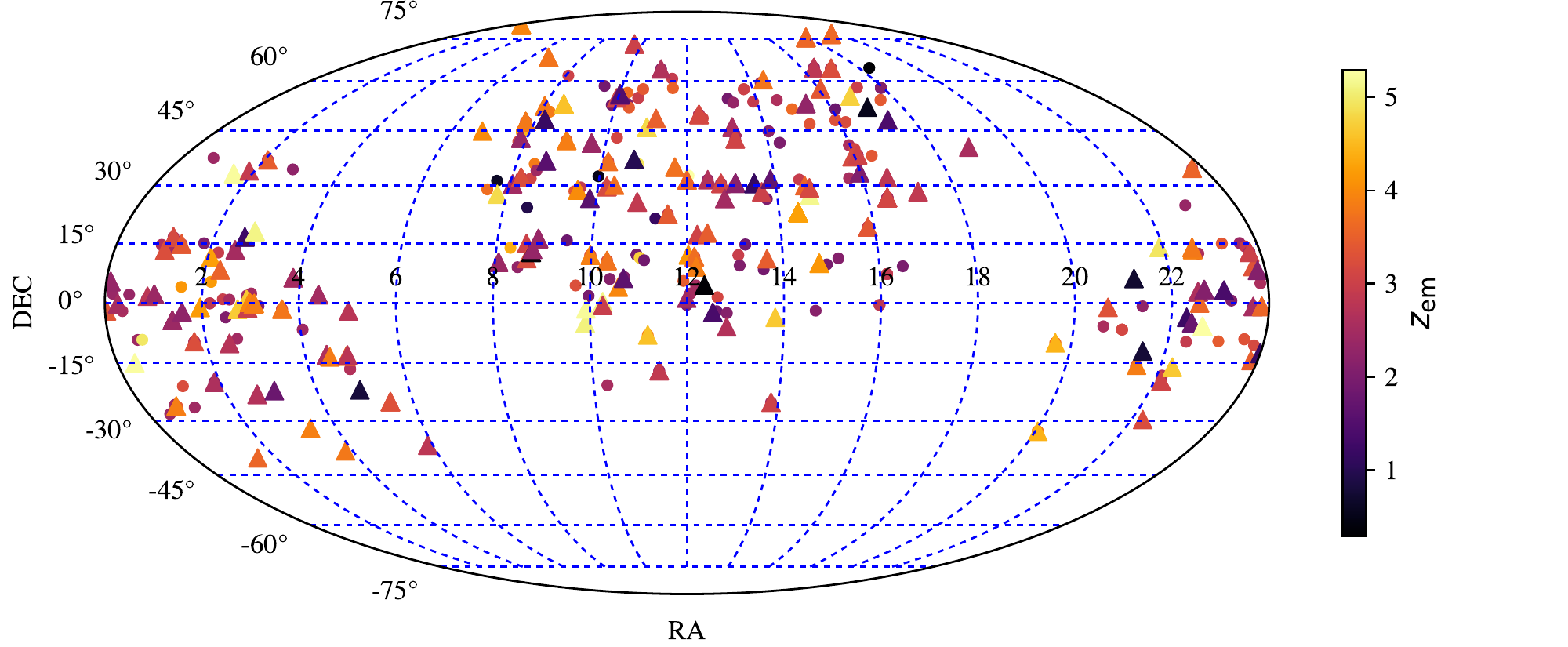}  
\caption{Distribution on the Sky of the KODIAQ DR2 quasars. Triangles
  correspond to new quasars as presented in Table 2, and circles
  representing quasars presented in DR1\label{fig_skyplot}}
\end{figure*}

Figure \ref{fig_zhist} shows the quasar redshift distribution for the
DR2 sample, including the distribution for the 130 new quasars alone.
DR2 spans a range in redshift from $0.07 < z_{\rm{em}} < 5.29$, with
a median redshift of $\bar{z}_{\rm{em}} = 2.586$ and a standard
deviation of $\sigma_{z_{\rm{em}}} = 0.918$.  The distribution of DR2
quasars on the sky is shown in Figure \ref{fig_skyplot}.
To illustrate the properties of DR2, we have performed a further
co-addition of the data to produce a single spectrum per quasar.  For
each quasar, the co-addition re-samples the data onto a single binning
and resolution. The figures in \cite{omeara15} largely summarize the DR1 in a different
manner, applying to the spectra co-added within a particular
instrument setup. 

\begin{figure*}[ht]
\epsscale{0.7} 
\plotone{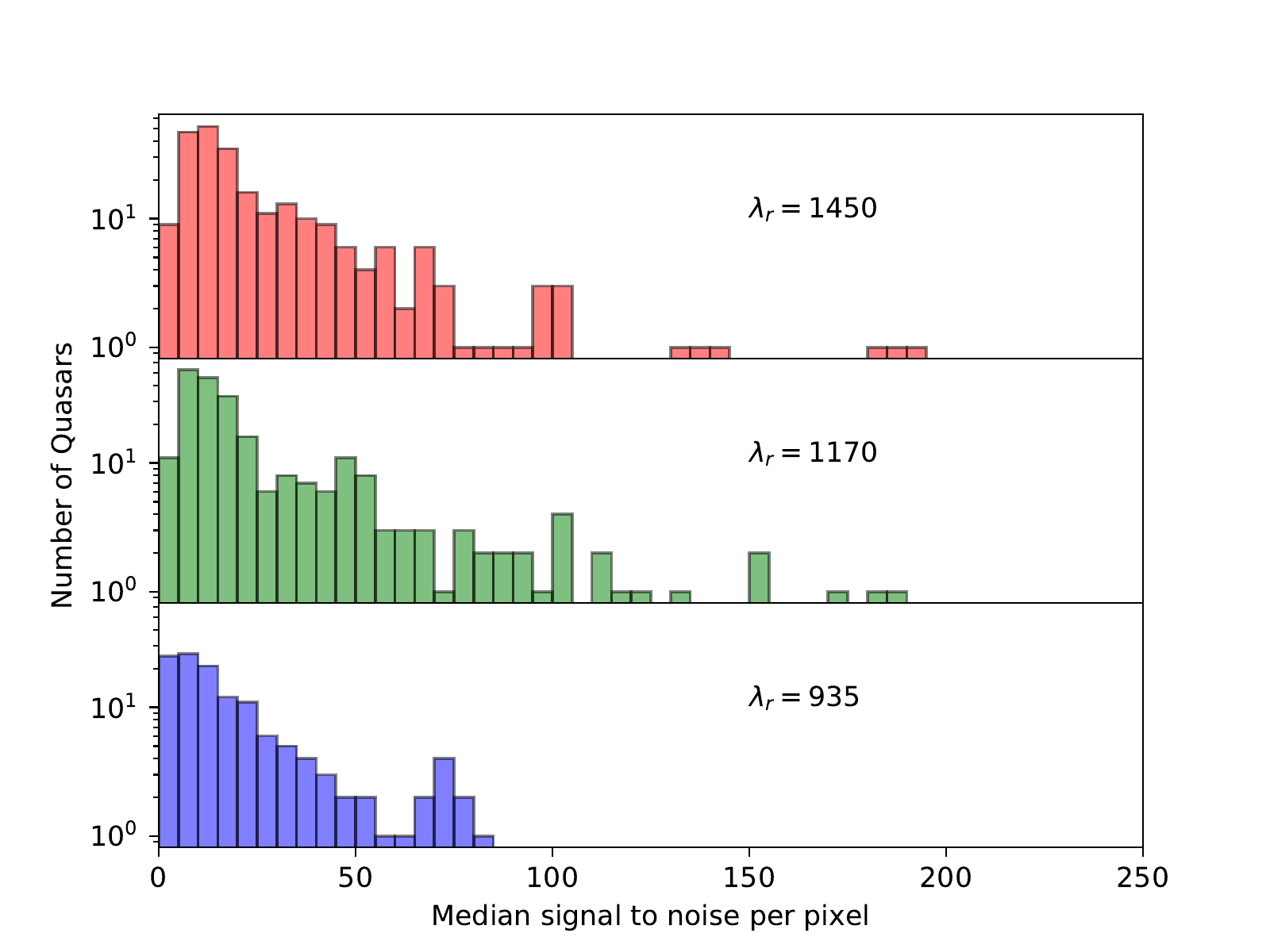}  
\caption{Signal to noise (per pixel) distribution of the co-added spectra in  the DR2 sample.\label{fig_snrhist}}
\end{figure*}

Figure \ref{fig_snrhist} displays the median S/N, as calculated by $\frac{S}{N} = \frac{1.0}{\sigma}$ per pixel
of the DR2 within $\pm 5$\AA\ of three rest wavelengths.  The pixels
are a constant 2.1 \kms\ for pre-2004 observations, 2.6 \kms\ for
post-2004 observations, or 2.6 \kms\ for co-additions of observations
mixing pre- and post-2004 data.  The rest
wavelengths were chosen to illustrate the viability for DR2 to be used
for surveys of metal line, \lya, and Lyman limit absorption.  As can
be seen in Figure \ref{fig_snrhist}, a significant number of quasars
have high S/N, with many exceeding S/N 50 per pixel. 

\subsection{Cosmological Properties}\label{cosmological}
The DR2 represents a significant increase over DR1 in its ability to
provide a rich data sample for cosmological studies.  We highlight a
number of the DR2 properties below with respect to its ability to be
used for surveys of metal line and HI absorption. 

\begin{figure*}[ht]
\epsscale{0.7} 
\plotone{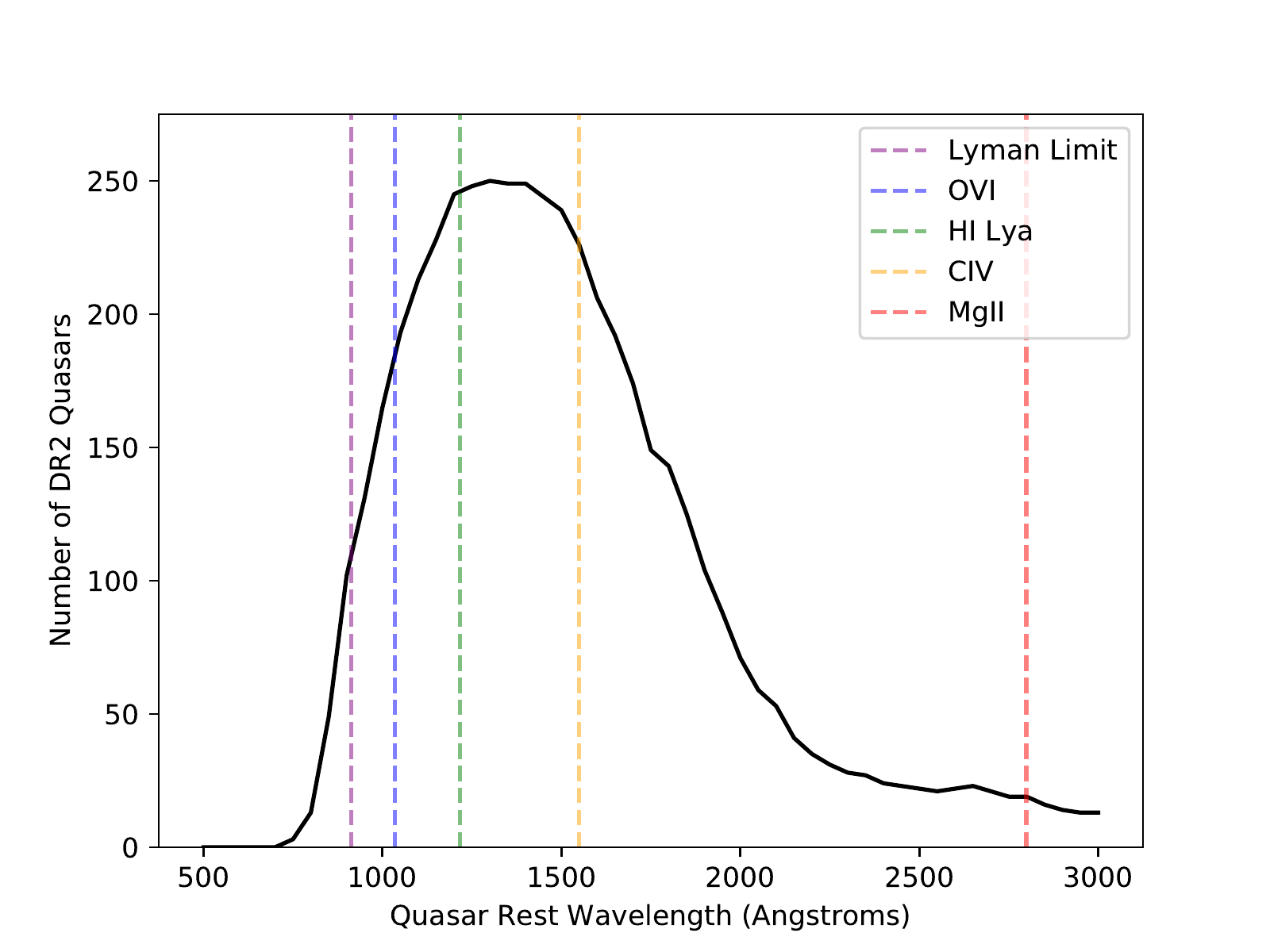}  
\caption{QSO redshift rest wavelength coverage of the DR2 sample. The
  vertical lines correspond to the rest wavelengths of various
  commonly studied ions.\label{fig_resthist}}
\end{figure*}

\begin{figure*}[ht]
\epsscale{0.7} 
\plotone{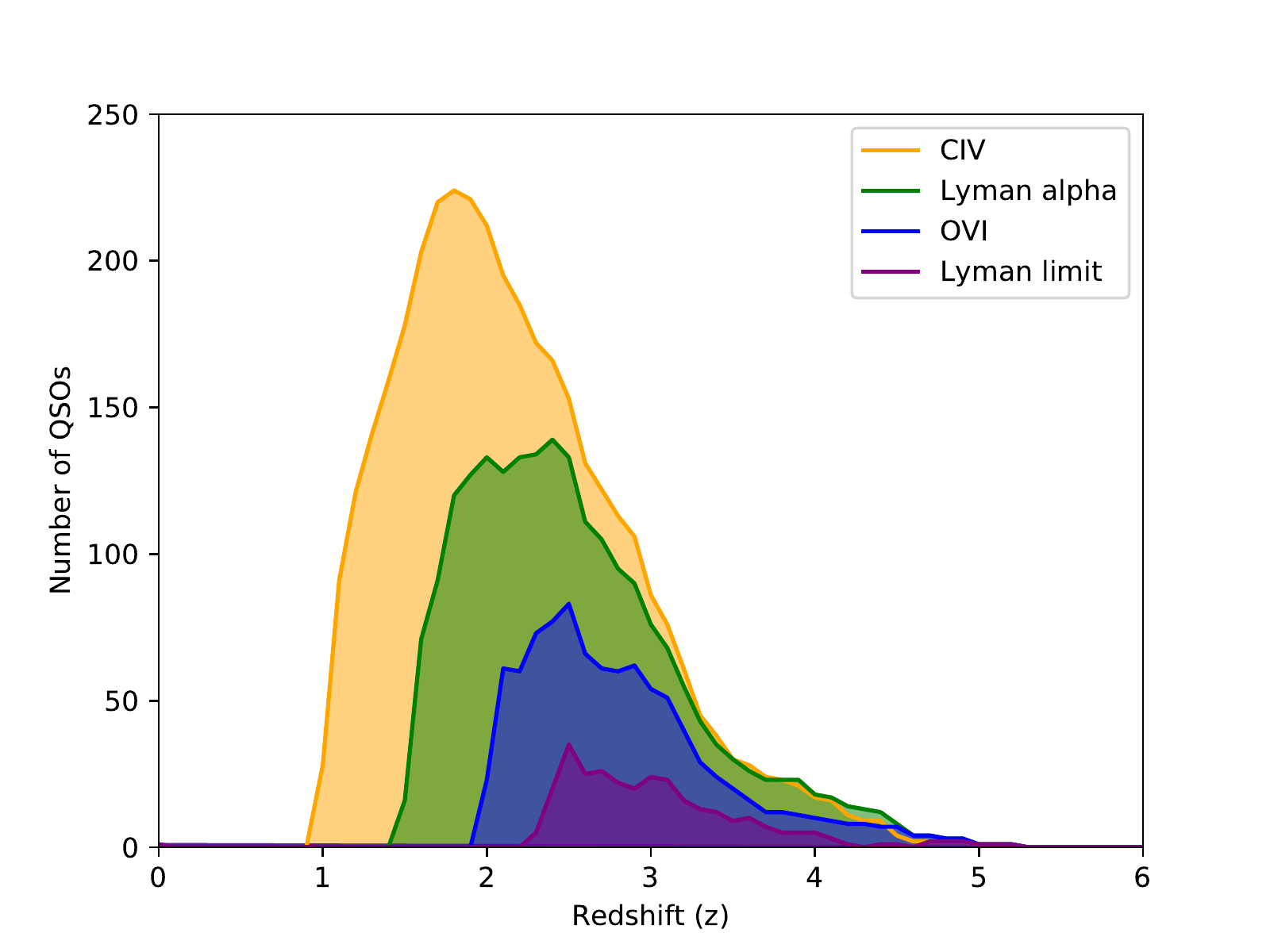}  
\caption{Redshift coverage of intervening absorption from various ions in the DR2 sample.\label{fig_ions}}
\end{figure*}

Figure \ref{fig_resthist} shows the rest wavelength (quasar redshift frame)
spectral coverage of DR2.  Well over 200 quasars in DR2 cover rest
wavelengths corresponding to \lya\ and C~IV, and approximately 100
quasars cover the quasar Lyman limit.  Figure \ref{fig_ions} displays
the redshift range over which certain ions may appear in the DR2
spectra.  It bears noting that this figure is optimistic in that
although certain ions (such as C~IV) could appear in DR2 spectra, they
may be blended with \lyaf\ absorption from higher redshifts.
Nevertheless, Figure \ref{fig_ions} demonstrates that searches for
ionic absorption in DR2 data, particularly at redshifts near $z=2.5$ are likely to have strong statistical power.  

\begin{figure*}[ht]
\epsscale{0.7} 
\plotone{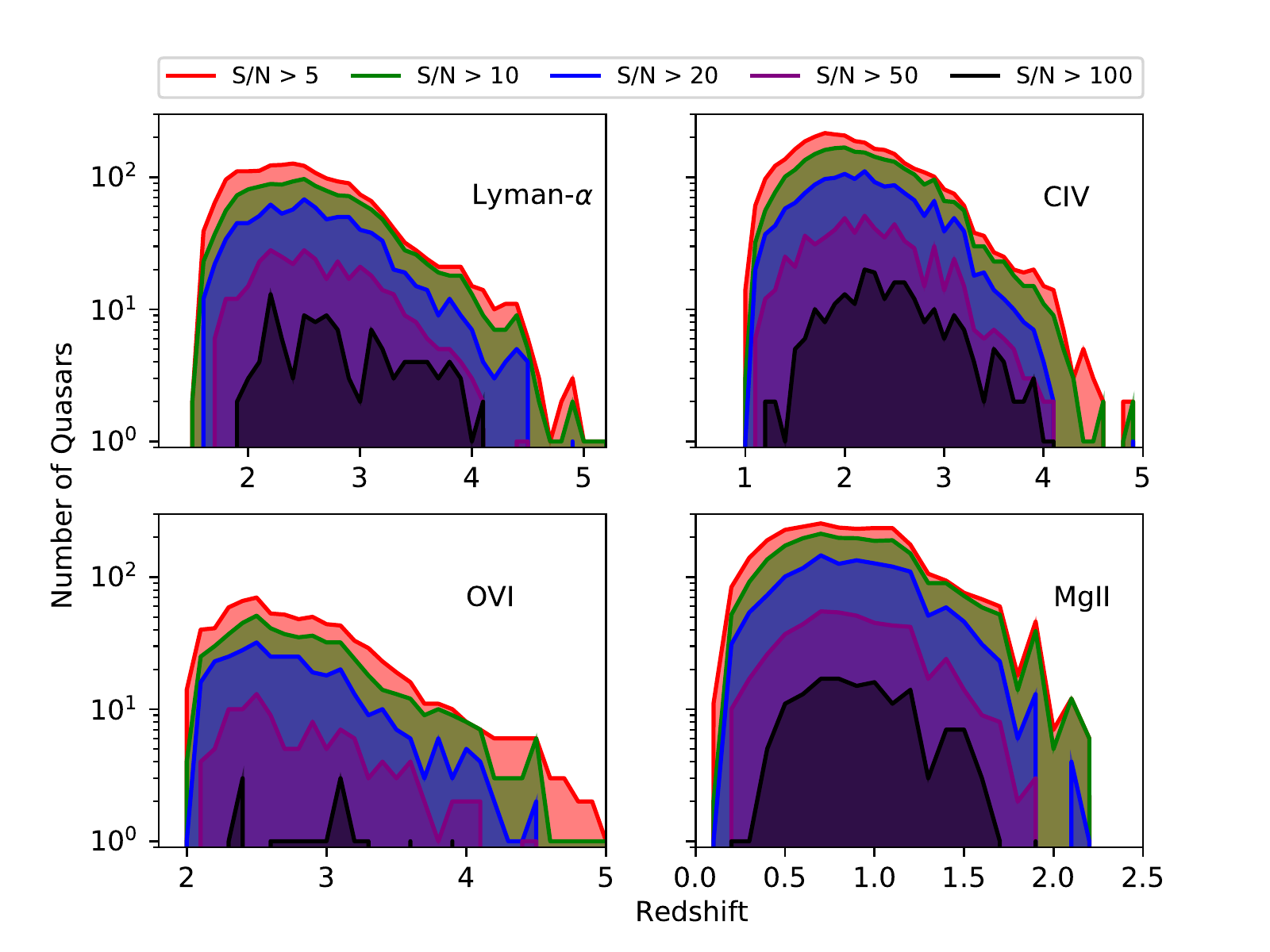}  
\caption{Signal to noise  distribution versus redshift for key ions in the DR2 sample.\label{fig_snrions}}
\end{figure*}

A further illustration of this point
is found in figure \ref{fig_snrions} which displays the redshift coverage
of certain ions and their numbers at or above certain S/N cuts in the
data.  Again, \lya\ and \ion{C}{4} stand out in terms of their numbers and
data quality, but we also note that a significant number of high
quality spectra exist covering \ion{Mg}{2} absorption, albeit subject to the same caveat that some of the \ion{Mg}{2} spectral coverage overlaps with higher redshift \lyaf.  Many tens of spectra have S/N in excess of 100, making DR2 the largest collection of high resolution, high signal to noise quasar spectra openly available to the community.

\begin{figure*}[ht]
\epsscale{0.7} 
\plotone{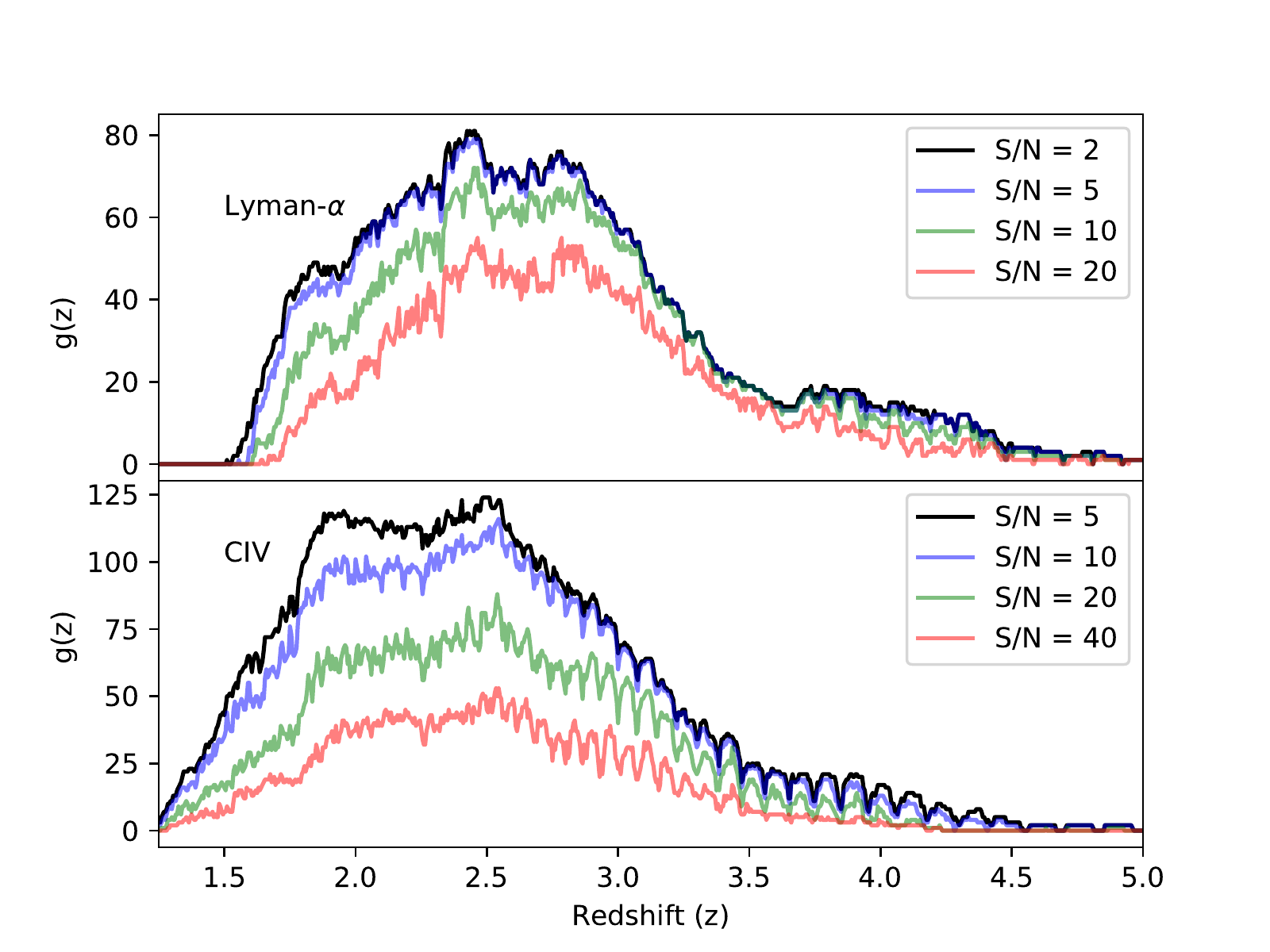}  
\caption{Redshift sensitivity function $g(z)$ for \lya\ and \ion{C}{4} in
  the DR2 as a function of S/N ratio.\label{fig_goz}}
\end{figure*}

Finally, Figure \ref{fig_goz} gives the redshift sensitivity function $g(z)$
for DR2.  Unlike the previous figures, we consider specific wavelength
ranges in the spectra to calculate $g(z)$.  As in \cite{omeara15}, we
select only the regions within a given quasar spectrum between 3000
\kms\ to the red of the quasar \lyb\ line and 3000 \kms\ to the blue
of the quasar \lya\ line for the $g(z)$ calculation of the redshift
sensitivity to \lya\ absorption.  
For \ion{C}{4}, we use the same 3000 \kms\
windows, but for the \lya\ and \ion{C}{4} emission lines, respectively.  The
high frequency variations in $g(z)$ stem from the blaze function of
the echelle orders, and various sharp features usually arise from
detector defects or gaps in spectral coverage.  

\section{Summary and Future}\label{future}
We have presented here and made public at the KOA the DR2 of the
KODIAQ survey.  DR2 contains spectra of 300 individual quasars
obtained with HIRES since 1995.  The data vary in signal to noise and
resolution, but a significant subset of the DR2 is of high enough
quality to facilitate a number of precision studies of the
intergalactic and circumgalactic medium between $0.15 < z < 5$.

With DR2, a significant fraction of the total number of
quasar observations made with HIRES since 1995 has been released to
the community.  We intend one final data release of HIRES data which
will mix the remaining pre- and post-2004 data available from the KOA
that can be nominally reduced.  This final data release is anticipated
in 2019--2020.  As part of KODIAQ z, we will also reduce and distribute all
quasars observed with the ESI instrument on Keck-II in $\sim 2020$.

When using data products from DR2, in addition to the standard KOA acknowledgement 
(including acknowledgement to the original PIs of each program), 
we request that the community please include the following acknowledgement: 
``{\it Some/all the data presented in this work were obtained from the 
Keck Observatory Database of Ionized Absorbers toward QSOs (KODIAQ), 
which was funded through NASA ADAP grants NNX10AE84G and NNX16AF52G  along
with NSF award number 1516777}'' 
and to cite this paper and \citet{omeara15}. 

\acknowledgements
Support for this work was made by NASA through the Astrophysics
Data Analysis Program (ADAP) grants NNX10AE84G and NNX16AF52G, along
with NSF grant award number 1516777.
This research has made use of the Keck Observatory Archive (KOA),
which is operated by the W. M. Keck Observatory and the NASA Exoplanet Science 
Institute (NExScI), under contract with the National Aeronautics and 
Space Administration.   The data presented herein were obtained at the
W.M. Keck Observatory, 
which is operated as a scientific partnership among the California
Institute of Technology, 
the University of California and the National Aeronautics and Space
Administration. 
The Observatory was made possible by the generous financial support of
the W.M. Keck Foundation.

The authors wish to recognize and acknowledge the very significant
cultural role and reverence that the summit of Mauna Kea has always had within 
the indigenous Hawaiian community.  We are most fortunate to have the 
opportunity to conduct observations from this mountain.
The authors wish to recognize and sincerely appreciate 
the work of the entire WMKO
staff over the last two-plus decades, and to the efforts of the team at the
NASA Exoplanet Science Institute (NExScI) who are responsible for maintaining
the KOA.  Finally, we continue to dedicate this work to the late astronomers
Arthur Wolfe and Wal Sargent, and extend the dedication to the late Jerry
Nelson, without whom the Keck telescopes and their great impact on
science would not exist.

\begin{deluxetable}{ccc}
\tablewidth{0pc}
\tablecaption{HIRES Deckers used in KODIAQ DR1}

\tablehead{
\colhead{Decker} &
\colhead{Width} &
\colhead{Spectra Resolution}\\
 \colhead{}& \colhead{(arcsec)} & \colhead{(FWHM)}
}

\startdata
B2 & 0.574 & $72,000$\\
B5 & 0.861 & $48,000$\\
C1 & 0.861 & $48,000$\\
C2 & 0.861 & $48,000$\\
C5 & 1.148 & $36,000$\\
D1 & 1.148 & $36,000$\\
E3 & 0.400 & $103,000$
\enddata
\end{deluxetable}

\newpage

\begin{deluxetable}{llllccccl}

\tablewidth{0pc}
\tablecaption{The KODIAQ DR1 Sample}
\tabletypesize{\tiny}

\tablehead{
\colhead{Object} &
\colhead{R.A.$^a$} &
\colhead{Dec.$^a$} &
\colhead{$z_{\rm em}$$^a$} &
\colhead{Observation} &
\colhead{PI} &
\colhead{Total Exp,} &
\colhead{Decker} &
\colhead{Wavelength}
\\
\colhead{} & \colhead{(J2000)} & \colhead{(J2000)} &  \colhead{} & \colhead{Date}  & \colhead{}& \colhead{Time (s)} &\colhead{} & \colhead{Coverage (\AA)}
}

\startdata
J000150$-$015940 & 00:01:50.0  & $-$01:59:40    & 2.810 & Sep 1997 & Wolfe & 25000 & C5 & 4111$-$6520\\

J000520+052410 & 00:05:20.21 & +05:24:10.79 & 1.887 & Oct 1999 & Sargent & 8000 & C1 & 3307$-$4849\\
\nodata  & \nodata  & \nodata  & \nodata  & Oct 1999 & Sargent & 8000 & C1 & 3340$-$4888\\
\nodata  & \nodata  & \nodata  & \nodata  & Sep 1998 & Steidel & 3600 & C1 & 3217$-$4723\\

J001602$-$001224 & 00:16:02.40 & $-$00:12:24.9  & 2.087 & Nov 1996 & Roth & 9000 & C1 & 3766$-$6191\\
\nodata  & \nodata  & \nodata  & \nodata  & Nov 1996 & Roth & 5400 & C1 & 3903$-$6875\\

J001708+813508 & 00:17:08.47 & +81:35:08.13 & 3.366 & Aug 1995 & Cowie & 3600 & C1 & 4414$-$6873\\
\nodata  & \nodata  & \nodata  & \nodata  & Aug 1995 & Cowie & 5400 & C1 & 4437$-$6911\\
\enddata
\tablecomments{$^a$J2000 coordinates and quasar redshifts are
  determined by passing the quasar coordinates from the raw data
  header through SIMBAD or SDSS and choosing the appropriate match.}
\end{deluxetable}

\newpage

\begin{deluxetable}{lcccc}
\tablewidth{0pc}
\tablecaption{The KODIAQ DR2 Sample}
\tabletypesize{\tiny}
\tablehead{
\colhead{Object} &
\colhead{$z_{\rm em}$} &
\colhead{Total Exp,} &
\colhead{Wavelength} &
\colhead{Rest Wavelength}
\\
\colhead{} &   \colhead{} &  \colhead{Time (s)}  &\colhead{Coverage (\AA)} &\colhead{Coverage (\AA)$^b$}
}
\startdata
J000150-015940 & 2.81 & 25000 & 4111$-$6520&1079$-$1711\\
J000520+052410 & 1.887 & 19600 & 3217$-$4888&1114$-$1693\\
J000931+021707 & 2.35 & 1200 & 3334$-$6198&995$-$1850\\
J001602-001224 & 2.087 & 14400 & 3766$-$6875&1219$-$2227\\
J001708+813508 & 3.366 & 19800 & 4414$-$7791&1010$-$1784\\
J002127-020333 & 2.596 & 700 & 3366$-$6198&936$-$1723\\
J002208-150539 & 4.528 & 13500 & 5293$-$7792&957$-$1409\\
J002830-281704 & 2.4 & 600 & 3366$-$6198&989$-$1822\\
\enddata
\tablecomments{$^b$ Rest wavelength calculated at the Quasar redshift}
\end{deluxetable}

\newpage

\end{document}